\documentclass[aps,prd,reprint,groupedaddress,onecolumn,showpacs]{revtex4-1}
\usepackage{color}
\usepackage{graphicx}
\usepackage{bm}
\usepackage{amssymb,amsmath,amsfonts, mathrsfs}

\usepackage{url}
\begin{document}

\begin{flushright}
USTC-ICTS-16-13
\end{flushright}
\title {Virtual states and generalized completeness relation in the Friedrichs Model}
\author{Zhiguang Xiao}
\email[]{xiaozg@ustc.edu.cn}
\affiliation{Interdisciplinary Center for Theoretical Study, University of Science
and Technology of China, Hefei, Anhui 230026, China}

\author{Zhi-Yong Zhou}
\email[]{zhouzhy@seu.edu.cn}
\affiliation{Department of Physics, Southeast University, Nanjing 211189,
P.~R.~China}
\affiliation{
Kavli Institute for Theoretical Physics China, CAS, Beijing 100190, China}


\date{\today}

\begin{abstract}
We study the well-known Friedrichs model, in which a discrete state is
coupled to a continuum state. By examining the pole behaviors of the
Friedrichs model in a specific form factor thoroughly, we find
that, in general, when the bare discrete state is below the threshold
of the continuum state, there should also be a virtual-state pole
accompanying the bound-state pole originating from the bare discrete
state as the coupling is turned on. There are also other
second-sheet poles originating from the singularities of the form
factor. We give a general argument for the existence of these two kinds
of states. As the coupling is increased to a certain value, the
second-sheet poles may merge and become higher-order poles. We then
discuss the completeness relations incorporating bound states, virtual
states, and resonant states corresponding to higher-order poles.

\end{abstract}


\maketitle

\section{Introduction}

Unstable states in quantum physics appear in a lot of fields in
modern physics, such as unstable nuclei in  nuclear physics and
resonances in particle physics. In  hadron physics especially,
unstable resonances always arise in the strong interactions, and more
and more newly observed resonant states are quoted in the Particle
Data Group Table~\cite{Agashe:2014kda}. However, many states  fall
outside the expectations  of
the conventional quark models such as the Godfrey-Isgur model
\cite{Godfrey:1985xj}, in which mesons are 
regarded as the bound state of a quark and an antiquark in some 
potential induced by QCD.
Typically, the enigmatic $\sigma$ and $\kappa$ resonances in $\pi\pi$
and $\pi K$
scatterings~\cite{Xiao:2000kx,Zheng:2003rw} can hardly be accommodated in
the conventional quark model, because they are strongly coupled with
the continuum states. In the higher energy region, e.g. for the
quarkonium-like states near or above the open-flavor thresholds, such
as $X(3872)$, $D_0^*(2318)$, $D_{s0}^*(2317)$, $X(3900)$, the conventional quark
models neither
work well. An interesting approach by taking into account the hadron loop
effects provides a generally good description to the masses and widths
of the resonances ranging
from the light scalars~\cite{Zhou:2010ra} to the heavier
charmed, charmed-strange, and charmonium-like
spectra~\cite{Eichten:1978tg,Zhou:2011sp,Zhou:2013ada}.  From the
point of view of the hadron-loop model, one may roughly regard
 these resonances as being composed partly by the discrete spectra and
the continuum spectra of the free Hamiltonian without their interaction. How to describe these resonances in terms of
these components in a more rigorous way in quantum theory inspires us
to look at the mathematical description of the resonances.

The description of resonance, i.e. Gamow states, cannot be formulated in
the usual Hilbert space language, since it has complex energy
eigenvalue. In order to complete the task, one has to enlarge the
usual Hilbert space to a rigged Hilbert space (RHS). The main point is to have a Gel'fand
triplet $\Omega\subset \mathscr H \subset \Omega^\times$, where
$\mathscr H$ is the usual Hilbert space of the normalizable states,
$\Omega$ is a nuclear space which is dense in $\mathscr H$, and
$\Omega^\times$ is the space of the antilinear continuous functionals
on the nuclear space. Gamow states must be in the larger $\Omega^\times$,
since it is the generalized eigenstate of the full Hamiltonian with
complex eigenvalues. The descriptions of in-state and out-state are
using different rigged Hilbert spaces, $\Omega_{\pm}\subset \mathscr H
\subset \Omega_\pm^\times$ where the subscript "$-$" denotes the out-state
space and $+$ denotes the in-state space. The triplet of state spaces
can be mapped to the complex function spaces $D_{\mp}\subset \mathscr
H^2_\mp\subset D_{\mp}^\times $ where $D_{\mp}=S\cap \mathscr
H^2_{\mp}|_{\mathbb R^+}$, respectively, to form a representation,
where $S$
is the Schwartz space,  $\mathscr H^2_{\mp}$ is the so-called Hardy
space in which the functions are analytic on  $\mathbb C_\mp$, and
$|_{\mathbb R^+}$ means restriction on $\mathbb R^+$. There are also
two kinds of Gamow states, $|z_R^-\rangle\in \Omega_-^\times$,
$|z_R^+\rangle\in \Omega_+^\times$ denoting the decaying state and
growing states, which correspond to the lower and upper second-sheet
poles of the $S$ matrix, respectively. For further detailed discussion
on the mathematical foundation, the readers are referred to~\cite{Bohm:1989,Gadella:2004}.

The Friedrichs model~\cite{Friedrichs:1948} is  a
solvable model which demonstrates the property of the Gamow state. In the simplest Friedrichs
model~\cite{Friedrichs:1948} there are a discrete eigenstate and a
continuum eigenstate of the free Hamiltonian, which couple to
each other through an interaction term in the full Hamiltonian.  The
eigenstates of the full Hamiltonian can be worked out exactly. If the
free discrete state is located above the threshold of the continuum,
it will become unstable Gamow states represented as a pair of  poles on the second Riemann sheet of
the $S$ matrix. The pole on the lower second sheet corresponds to the
decaying Gamow state and the one on the upper corresponds to the
growing state. The Gamow state can be explicitly written down as a
vector in the RHS as a linear combination of the free
discrete state and the continuum states.  Besides these resonance
poles generated from the discrete states, there could  also be other
poles introduced by the form factors. This was noticed  in 
Ref.\cite{Likhoded:1997} by studying some special form factors
and the authors argue that  it is a possible origin of extra
states in particle physics. In present paper, we will give a
general arguement of the existence of these states introduced by the
form factors. If the free discrete state is
below the threshold of the continuum, there is a bound-state pole
on the physical sheet below the threshold after turning on the interaction. However, people usually do not notice
that,  besides this bound-state pole, there will also be a virtual state  on the unphysical sheet
originating from the same discrete state. This virtual state pole is the so-called shadow pole proposed by Eden~\cite{Eden:1964zz} in $S$-matrix
theory.  In our present paper, we will demonstrate the
existence of this virtual-state pole in  the Friedrichs model by
studying some example form factors and give a
general argument of its existence.  We also track the pole
trajectories as the couplings change  continuously and we find that even
for such a simple form factor there are unexpected pole structures.
The pair of the resonance poles can merge to be a degenerate double
pole and then separate on the real axis, becoming two virtual-state poles. One of
these virtual poles may meet with the virtual pole generated by the
form factor and then they separate into the complex plane becoming a pair
of resonance poles. In some special critical condition, the pair of
resonance poles and the virtual pole originating from the form factor
may meet on the negative real axis and form a triple pole.  The double pole
can be regarded as two degenerate states and the triple pole as
three degenerate ones. In fact, the higher order degenerate states
are not
the eigenstates of the Hamiltonian. In the bases of these degenerate
states, the Hamiltonian is represented as the matrix of Jordan
form. One would wonder whether these states would contribute to the
completeness relations. In fact, for nondegenerate Gamow states for the
resonances, by suitably incorporating the integral contour information
in the
continuum state as a distribution, the identity operator can be decomposed in the continuum
states and the Gamow states~\cite{Prigogine:1991}. Here  we will
generalize the completeness relations to
the degenerate higher order states in the Friedrichs model.

The structure of this paper is as follows: Section
\ref{sect:Friedrichs} introduces the Friedrichs model. In 
Sec.~\ref{sect:Examples}, we examine pole trajectories of the Friedrichs
model with a kind of example form factor. In Sec.~\ref{sect:General}, we give a general argument for the existence of the
virtual state accompanying the bound state and the states generated by
form factor.   In Sec.
\ref{sect:Complete}, we study the completeness relations with higher-order
degenerate states. Section \ref{sect:Conclude} is the conclusion
and discussion.

\section{\label{sect:Friedrichs} Introduction to the Friedrichs model}
The simplest Friedrichs model~\cite{Friedrichs:1948} includes a free
Hamiltonian $H_0$ with a simple continuous spectrum, which is
$\mathbb{R}^+\equiv [0,\infty)$, plus a discrete eigenvalue
$\omega_0$ imbedded in this continuous spectrum~($\omega_0>0$). An
interaction $V$ between the continuous and discrete parts is produced
so that the discrete state of $H_0$ is dissolved in the continuous
spectrum and a resonance is produced. In fact, all the following
solutions also apply to $\omega_0<0$ cases, so we would not restrict
the domain of $\omega_0$ for the moment. We denote  the discrete state
of $H_0$ by $|1\rangle$ and the continuum state by the $|\omega\rangle$, that
is,
\begin{eqnarray}
H_0|1\rangle=\omega_0|1\rangle,\nonumber\\
H_0|\omega\rangle=\omega|\omega\rangle.
\end{eqnarray}
The free Hamiltonian is then
\begin{eqnarray}
H_0=\omega_0|1\rangle\langle 1|+\int_0^\infty \omega|\omega\rangle\langle\omega |\mathrm{d}\omega,
\end{eqnarray}
and the interaction $V$ is written as
\begin{eqnarray}
V=\lambda\int_0^\infty [f(\omega)|\omega\rangle\langle 1
|+f(\omega)|1\rangle\langle \omega |]\mathrm{d}\omega\,.
\end{eqnarray}
 For simplification, we assume that $f(\omega)$ is real for above the threshold, i.e. $\omega>0$. In general, it could be a complex function. The normalizations and orthogonal conditions for the free states are
\begin{eqnarray}
\langle 1|1\rangle=1,\langle 1|\omega\rangle=\langle \omega|1\rangle=0,\langle\omega|\omega'\rangle=\langle\omega'|\omega\rangle=\delta(\omega-\omega').
\end{eqnarray}
We will solve the eigenstate $|\Psi(x)\rangle$ of $H=H_0+V$ with eigenvalue
$x$,
\begin{eqnarray}\label{eigenf}
H\Psi(x)=x|\Psi(x)\rangle.
\end{eqnarray}
Since $|1\rangle$ and $|\omega\rangle$ form a complete set, the
eigenstate  $|\Psi(x)\rangle$ can be expressed  in terms of $|1\rangle$ and $|\omega\rangle$,
\begin{eqnarray}
|\Psi(x)\rangle=\alpha(x)|1\rangle+\int_0^\infty\psi(x,\omega)|\omega\rangle \mathrm{d}\omega.
\label{eq:EigenV}
\end{eqnarray}
Note that here $|\Psi\rangle$ is a vector in $\Phi^\times$, and it
only make senses as an antilinear functional on the vector
$|\phi\rangle\in \Phi$, $\langle \phi|\Psi\rangle$. So,
$\psi(x,\omega)$ should be treated as a distribution.
Substituting (\ref{eq:EigenV}) into Eq.(\ref{eigenf}), one can obtain the
following relations:
\begin{eqnarray}
(\omega_0-x)\alpha(x)+\lambda\int_0^\infty
f(\omega)\psi(x,\omega)\mathrm{d}\omega&=& 0,\nonumber\\
(\omega-x)\psi(x,\omega)+\lambda f(\omega)\alpha(x)&=& 0.
\end{eqnarray}
 Then, for real $x>0$, we have
\begin{eqnarray}
&\psi_\pm(x,\omega)=-\frac{\lambda\alpha(x)f(\omega)}{\omega-x\pm
i\epsilon}+\gamma(\omega)\delta(\omega-x)\,,
\nonumber\\
&(\omega_0-x)\alpha_\pm(x)+\lambda
f(x)\gamma(x)-\alpha_\pm(x)\lambda^2\int_0^\infty\frac{|f(\omega)|^2}{\omega-x\pm
i\epsilon}\mathrm{d}\omega=0.
\label{eq:two-rela}
\end{eqnarray}
where $\gamma(\omega)$ is an arbitrary function to be determined by
normalization. We have added the $\pm i\epsilon$ in order to define
the integral contour. Now, one can define
\begin{align}
\eta^{\pm}(x)=x-\omega_0-\lambda^2\int_0^\infty\frac{|f(\omega)|^2}{x-\omega\pm
i \epsilon}\mathrm{d}\omega\,,\label{eq:eta-pm}
\end{align}
and analytically continue $\eta^{\pm}$ to the complex plane
$\eta(x)$, 
and $\eta^{+}$ and $\eta^{-}$ are the boundary functions of $\eta(x)$
on the upper rim and lower rim of the cut on the positive axis,
respectively. Since $\eta(x)$ is a real-analytic function, satisfying the Schwartz
reflection relation 
$\eta^*(z)=\eta(z^*)$,  the analytically continued form factor  $G(x)\equiv
|f(x)|^2 $, being proportional to the imaginary part of $\eta(x)$,
should be anti-real-analytic, that is, $G(x^*)=-G(x)^*$.
In usual physical applications, $\eta(x)$ is real below the threshold
and complex  with a positive imaginary part above the threshold. This
requires $G(x)$ to positive real function above
the threshold and imaginary below the threshold.

After choosing the  normalization such that
$\langle\Psi_\pm(x')|\Psi_\pm(x)\rangle=\delta(x'-x)$, one obtains the solution  
\begin{eqnarray}
|\Psi_\pm(x)\rangle&=&|x\rangle+\lambda\frac{f(x)}{\eta^\pm(x)}\Big[|1\rangle+\lambda\int_0^\infty\mathrm{d}\omega\frac{f(\omega)}{x-\omega\pm
i \epsilon}|\omega\rangle\Big]
\label{eq:continuum-state-0}
\end{eqnarray}
The two sets of states $|\Psi_+(x)\rangle$ and $|\Psi_-(x)\rangle$ can
be understood as in-states and out-states, i.e. $|\Psi_\mp(x)\rangle\in
\Omega_\mp^\times$.

For complex eigenvalue $x$ not on the positive real axis, Eq.
(\ref{eq:two-rela}) would not have the $\gamma(\omega)$ term, and there is
 no need for
$i\epsilon$ in the denominator. In order for  $\alpha(x)$ to be
nonzero, the eigenvalue must satisfy $\eta(x)=0$. Depending on the
positions of the solutions, the eigenstates may be categorized into  resonant
states, bound states, or virtual states. 
\begin{enumerate}
\item {\it Resonant states.}
If $\eta(x)=0$ has a pair of complex conjugate solutions $z_R\in \mathbb C_-$ and
$z_R^*\in \mathbb C_+$ on the second sheet, the right eigenstates for
eigenvalue $z_R$ and $z_R^*$ can be
expressed as
\begin{eqnarray}
|z_R\rangle=N_R\Big(|1\rangle+\lambda\int_0^\infty\mathrm{d}\omega\frac{f(\omega)}{[z_R-\omega]_+}|\omega\rangle\Big),\nonumber\\
|z_R^*\rangle=N_R^*\Big(|1\rangle+\lambda\int_0^\infty\mathrm{d}\omega\frac{f(\omega)}{[z_R^*-\omega]_-}|\omega\rangle\Big),
\label{eq:Gamow-state-right}
\end{eqnarray}
which are the Gamow states satisfying $H|z_R\rangle =z_R|z_R\rangle$
and $H|z^*_R\rangle=z_R^*|z^*_R\rangle$. $[\dots]_{\pm}$ denotes the
continuation from the upper rim of the cut to
the lower second sheet for ``$+$'' or from the lower rim to the upper second sheet
for ``$-$'', and hence
the deformation of the integration path as in Fig. 
\ref{fig:contour} for ``$+$'' and the opposite for ``$-$'' is needed.   
\begin{figure}
\begin{center}
\begin{center}\includegraphics[width=7cm]{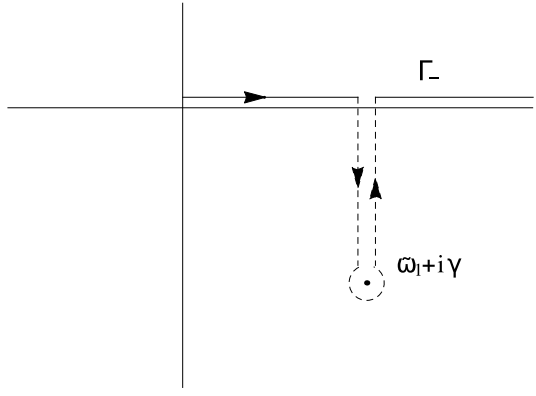}\end{center}
\end{center}
\caption{The deformation of the integral path. \label{fig:contour}}
\end{figure}

We have
also the left eigenstates, i.e. $\langle \tilde z_R| H=z_R\langle
\tilde z_R|$ and $\langle \tilde z^*_R| H=z^*_R\langle \tilde z^*_R|$,
\begin{eqnarray}
\langle \tilde
z_R|=N_R\Big(\langle1|+\lambda\int_0^\infty\mathrm{d}\omega\frac{f(\omega)}{[z_R-\omega]_+}\langle\omega|\Big),\nonumber\\
\langle
\tilde z_R^*|=N_R^*\Big(\langle1|+\lambda\int_0^\infty\mathrm{d}\omega\frac{f(\omega)}{[z_R^*-\omega]_-}\langle\omega|\Big).
\label{eq:Gamow-state-left}
\end{eqnarray}
The normalization is chosen as
$N_R=(\eta'^+(z_R))^{-1/2}=(1+\lambda^2\int d\omega
\frac{|f(\omega)|^2}{[(z_R-\omega)_+]^2})^{-1/2}$ such that $\langle
\tilde z_R| z_R\rangle =1$, since  the resonant state has zero norm
$\langle z_R|z_R\rangle=0$~\cite{Prigogine:1991, Nakanishi:1958}. 
\item {\it Bound states.}
If $\eta(x)=0$ have a solution on the
negative real axis on physical Riemann sheet, it  represents a bound
state.  The bound state with
eigenvalue $z_B$ can then be  represented as~\cite{Prigogine:2000,Prigogine:2001}
\begin{align}
|z_B\rangle=N_B
\Big(|1\rangle+\lambda\int_0^\infty\frac{f(\omega)}{z_B-\omega}|\omega\rangle\mathrm{d}\omega\Big)
\label{eq:Bound}
\end{align}
where $N_B=(\eta'(z_B))^{-1/2}=(1+\lambda^2\int d\omega
\frac{|f(\omega)|^2}{(z_B-\omega)^2})^{-1/2}$.

\item {\it Virtual states.}
If $\eta(x)=0$ has a solution on the negative real axis of the second
Riemann sheet, it corresponds to a virtual state.
The general discussion of the virtual states in the rigged Hilbert
space formulation can be found in Refs.~\cite{Gadella:1983vir} and~\cite{Gadella:1984vir}.  
For the simple virtual poles, similar to resonant states, there are
two kinds of states, $|z^+_v\rangle$ by analytical continuation from
the upper rim and
$|z^-_v\rangle$ lower rim of the cut to the second sheet
\begin{align}
|z^\pm_v\rangle=N^{\pm}_{v}
\Big(|1\rangle+\lambda\int_0^\infty\frac{f(\omega)}{[z_v-\omega]_\pm}|\omega\rangle\mathrm{d}\omega\Big),
\quad \langle \tilde z^\pm_v|= \langle  z^\mp_v| \,,
\label{eq:Virtual-state-right}
\end{align}
where $N^-_v=N^{+*}_v=(\eta'^+(z_v))^{-1/2}=(1+\lambda^2\int d\omega
\frac{|f(\omega)|^2}{[(z_v-\omega)_+]^2})^{-1/2}$.  
As discussed in Refs.~\cite{Gadella:1983vir} and~\cite{Gadella:1984vir},
$|z^-_v\rangle $ is only defined for $t>0$ representing the out-state
and $|z^+_v\rangle $ for $t<0$ representing the in-state.
In some accidental cases, there could be multiple zeroes for $\eta(x)$
on the negative real axis of the unphysical sheet,
corresponding to the degenerate virtual states, which will be discussed later.
\end{enumerate}

\section{\label{sect:Examples} Pole trajectories in the Friedichs model with
an example form factor}
In this section, we will use an example form factor to exhibit the
existence of these three kinds of states, and they can transform to
each other as the coupling changes.

As a simple integrable example, similar to~\cite{Likhoded:1997}, we
choose $|f(\omega)|^2=\frac
{\sqrt{\omega}}{\omega+\rho^2}$, $\rho>0$ and analyze the pole
structures for different parameters. The integral in $\eta(\omega)$ can be
worked out and analytically continued to the complex plane,
\begin{align*}
\eta(\omega)=\omega-\omega_0+\frac
{\pi\lambda^2}{\sqrt{-\omega}+\rho}=\omega-\omega_0+\frac
{\pi\lambda^2}{-i\sqrt{\omega}+\rho}\,.
\end{align*}
Choosing $-i$ for $\sqrt{-1}=(e^{-i\pi})^{1/2}$ in the denominator
makes the imaginary part of $\eta^+(x)$ positive at $x>0$.
The cut for $\omega$ is along the positive real axis.
Continued to the second sheet, $\eta$ reads
\begin{align*}
\eta^{II}(\omega)=\omega-\omega_0+\frac
{\pi\lambda^2}{-\sqrt{-\omega}+\rho}\,,
\end{align*}
where the superscript $II$ denotes the second Riemann sheet.

For future convenience, we turn to the momentum plane by making the change of the variable $\omega=u^2$.  The $u$ plane combines the
first sheet and the second sheet of the $\omega$ plane. The first sheet of
$\omega$ corresponds to the upper half-plane of $u$, and the second
sheet of $\omega$ corresponds to the lower half.  The equation
$\eta(u^2)=0$ can be recast into
\begin{align}
\frac 1{u+i\rho}(u^3+i\rho u^2-\omega_0u-i(\rho\omega_0-\pi\lambda^2))=0
\label{eq:eta-eq-0}
\end{align}
There are three solutions for this third-order algebraic equation.

{\it Case 1.} We first look at the case for $\omega_0>0$ and turning on the coupling
constant slowly such that $\omega_0>\frac
{\pi \lambda^2}{\rho}$. This case is studied in~\cite{Likhoded:1997} and we give a review
here for completeness and discuss more on the virtual states from
the form factor.  In this case, there is no bound state. The three
solutions correspond to a pair of resonant states and a virtual state.  If we set the three solutions
 to be $u_{1,2}=\pm\alpha -i\gamma$ and $u_3=-id$, the three solutions
can be expanded in orders of $\lambda$,
\begin{align}
E_{1,2}=&\omega_0-2d\gamma\mp 2i\gamma
\omega_0^{1/2}+O(\lambda^4)=\omega_0-\frac{\pi\lambda^2}{\rho\mp
i\omega_0^{1/2}}+O(\lambda^4)\,,
\\
E_3=&-(\rho-2\gamma)^2=-\rho^2+4\gamma \rho+O(\lambda^4)=-\rho^2+\frac
{2\rho\pi\lambda^2}{\omega_0+\rho^2}+O(\lambda^4)
\end{align}
The virtual-state pole at $E_3$ does not exist at exact $\lambda=0$,
 but as long as $\lambda$ runs away from $0$, the residue of the pole is not zero
and it appears near
the poles of the form factor on the second sheet.

The pair of resonance poles originates from the discrete state can be
represented as the Gamow states as in (\ref{eq:Gamow-state-right}). The
virtual state  can be represented as in (\ref{eq:Virtual-state-right}).
One may wonder whether, as $\lambda\to 0$, the virtual state
tends to $|1\rangle$ from (\ref{eq:Virtual-state-right}). This is not
correct. In fact, the integral term in (\ref{eq:Virtual-state-right})
is of $\mathcal O(\lambda^0)$, since at  $\lambda=0$,
  the integral is
divergent. In fact, for the example form factor, for small $\lambda$, the integral term should be of order $\lambda \mathcal
O(1/\lambda)\sim \mathcal O(\lambda^0)$. To see
this,  we take
the inner product $\langle \phi|z^-_v\rangle$, where $|z^-_v\rangle$
is defined in (\ref{eq:Virtual-state-right}) and
$|\phi\rangle \in
\Phi^-$ and $\langle \phi|\omega\rangle =\phi(\omega)\in S\cap
\mathscr H_-^2|_{\mathbb R^+}$. The
integral term in the inner product is
\begin{align}
\lambda\int_0^\infty\frac{\omega^{1/4}\phi(\omega)}{(\omega+\rho^2)^{1/2}[z_v-\omega]_+}\mathrm{d}\omega
=
\lambda\int_0^\infty\frac{\omega^{1/4}\phi(\omega)}{(\omega+\rho^2)^{1/2}(z_v-\omega)}\mathrm{d}\omega+2\pi
i\lambda\frac{z_v^{1/4}\phi(z_v)}{(z_v+\rho^2)^{1/2}}
\end{align}
The integral term in the first term is not singular and will
be $ \lambda\cdot\mathcal O(\lambda^0)\sim\mathcal O(\lambda^1)$. If we take
$z_v=e^{-i\pi}(\rho^2+a \lambda^2)$ on the second sheet, the second term
will be
\begin{align}
2\pi
i\lambda\frac{z_v^{1/4}\phi(z_v)}{(z_v+\rho^2)^{1/2}}=2\pi
i\frac{e^{-i\pi/4}\rho^{1/2}\phi(e^{-i\pi}\rho^2)}{(-a)^{1/2}}+O(\lambda^2)\sim
\mathcal O(\lambda^0)
\end{align}
So, the second term in (\ref{eq:Virtual-state-right}) is of the same
order as the first term.

\begin{figure}
\begin{center}\includegraphics[height=4cm]{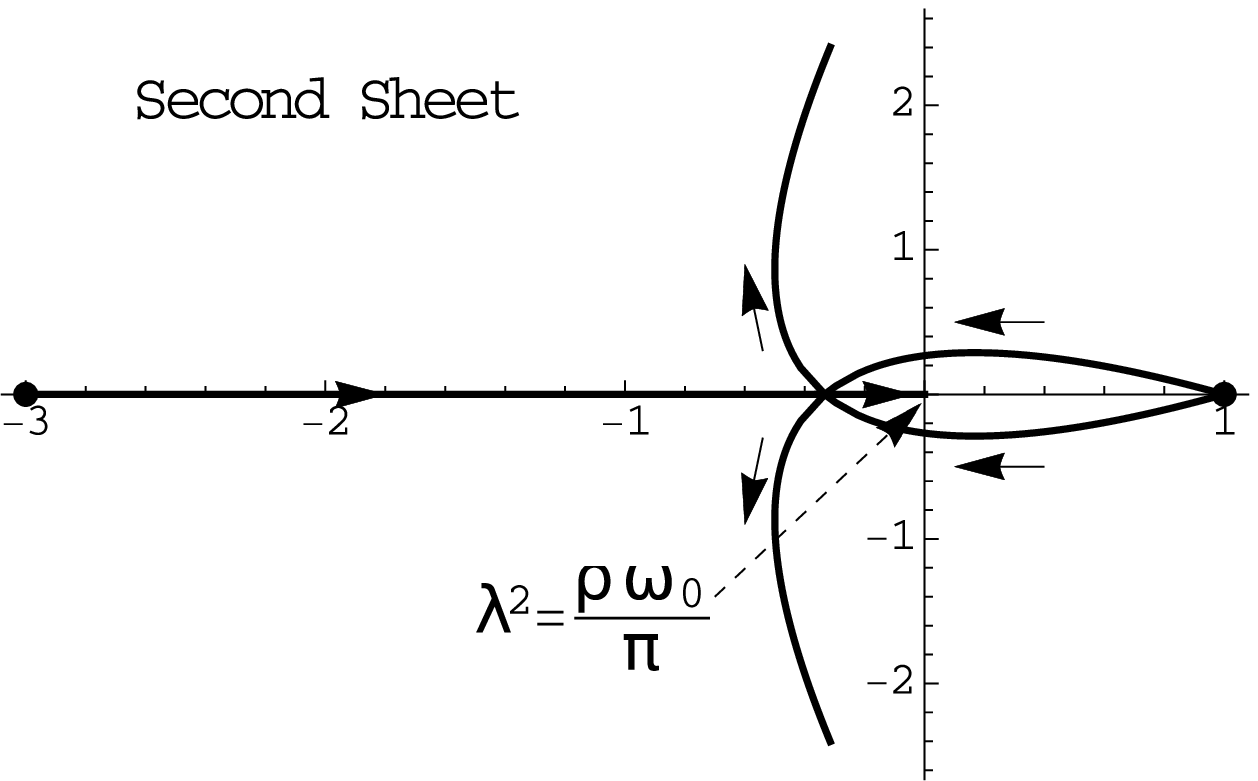}
\includegraphics[height=4cm]{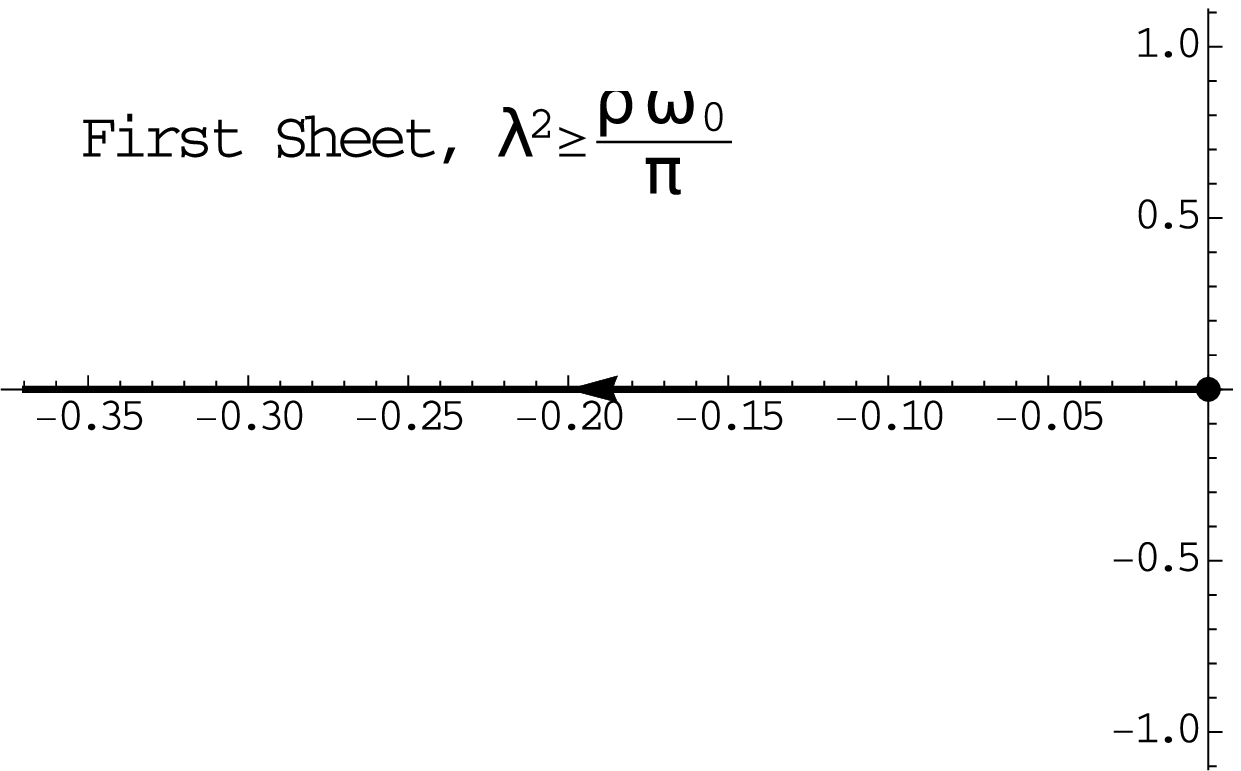}
\end{center}
\caption{The pole trajectories as $\lambda$ increases from $0$ when
$\omega_0=\frac 1 3 \rho^2$. $\rho^2=3,\omega_0=1$. The three second-sheet poles merge
and then separate. Two of them become a pair of resonance poles and the other
moves up to first sheet.}
\label{fig:triple}
\end{figure}

{\it Case 2. } For larger $\lambda$ satisfying
$0<\omega_0<\frac{\pi\lambda^2}{\rho}$, there are two cases: the two
second-sheet poles may or may not go back to the negative real axis to
form virtual states. The critical case separating these two cases is
the one in which the three poles come together on the negative axis
forming a third-order pole and then get away. Let us examine this
critical case first. We set the three degenerate poles to be
$u_{1,2,3}=-id$. One can find out  that only when
\begin{align}
d=\frac 1 3 \rho\,, \quad
\omega_0=\frac 1 3\rho^2
\,,\quad
\rho=\frac 3 2 (\pi\lambda^2)^{1/3}
\label{eq:triple-cond}
\end{align}
can  the three poles
merge to be a third-order pole, which means the degeneracy of the
three states (see Fig. \ref{fig:triple}). Then as
$\lambda$ becomes larger, two of them
run into the complex plane to be a pair of resonance poles, and the
other remains a virtual-state pole and goes up towards threshold. When
$\lambda^2> \frac{\rho\omega_0}\pi$, the virtual state will go through
the threshold to the first sheet, becoming a bound state and moving downward towards
minus infinity.

According to the discussion in~\cite{Gadella:1997}, the triple pole
can be expressed as three kinds of  states: the first is the same as the
ordinary virtual state as in (\ref{eq:Virtual-state-right}) and the
other two come from the higher-order residues of the continuum state
\begin{align}
| z^\pm_{v2}\rangle=-N_{v2}
\lambda\int_0^\infty\frac{f(\omega)}{([z_v-\omega]_{\pm})^2}|\omega\rangle\mathrm{d}\omega
,
\quad \langle \tilde z^\pm_{v2}|= \langle  z^\mp_{v2}| \,,
\\
| z^\pm_{v3}\rangle=N_{v3}
\lambda\int_0^\infty\frac{f(\omega)}{([z_v-\omega]_{\pm})^3}|\omega\rangle\mathrm{d}\omega\,,
\quad \langle \tilde z^\pm_{v3}|= \langle  z^\mp_{v3}| \,.
\end{align}
The normalization for the simple virtual state cannot be used here, since
$\eta(z)$ has a third-order zero, $\eta'(z_v)=0$ and also
$\eta''(z_v)=0$. We can choose $\langle \tilde
z_{v3}^\pm|z_{v}^\pm\rangle=1$ and $\langle\tilde z_{v2}^\pm|z_{v2}^\pm
\rangle =1$. Then $N_v=N_{v_2}=N_{v3}=(6/\eta''')^{1/2}$.
The Hamiltonian is not diagonalized by these states and can only be
represented as Jordan form, that is,
\begin{align}
H|z^\pm_{v3}\rangle=z_v |z^\pm_{v3}\rangle+ 2|z^\pm_{v2}\rangle\,,
\\
 H|z^\pm_{v2}\rangle=z_v |z^\pm_{v2}\rangle+ |z^\pm_{v}\rangle\,.
\end{align}

\begin{figure}
\includegraphics[height=4cm]{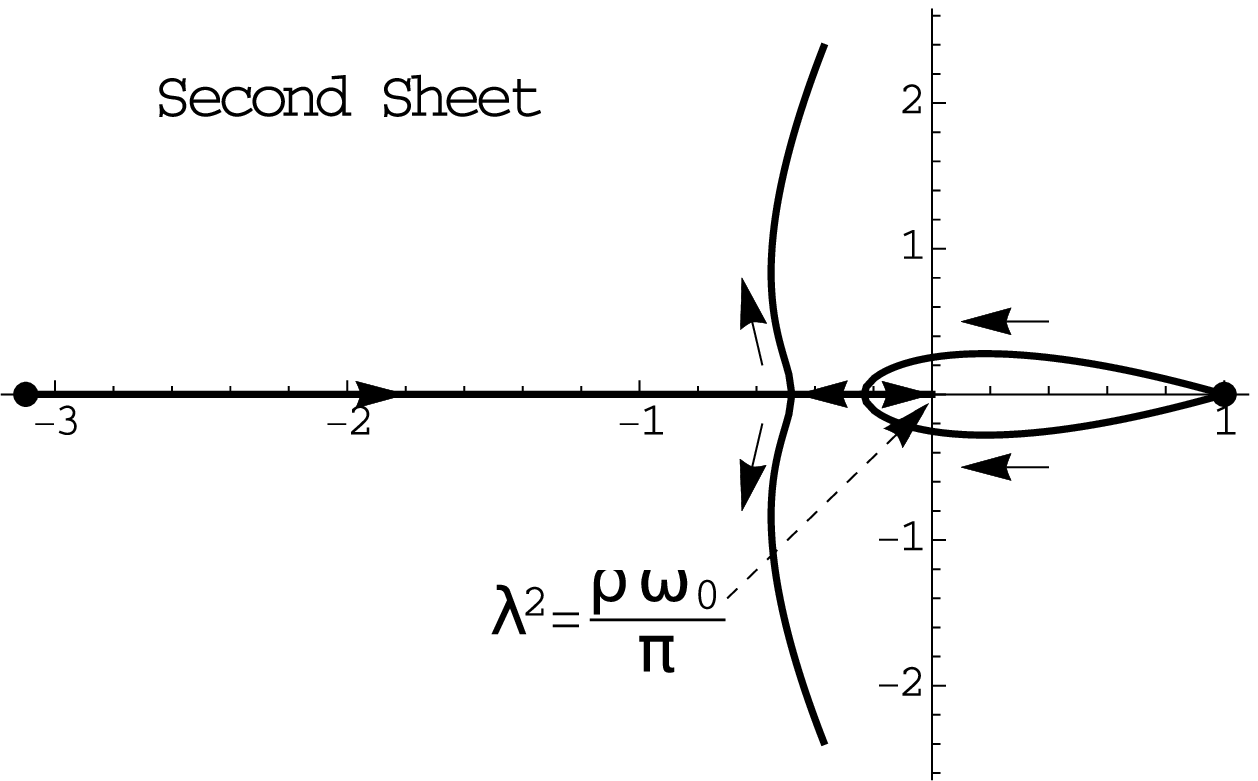}
\includegraphics[height=4cm]{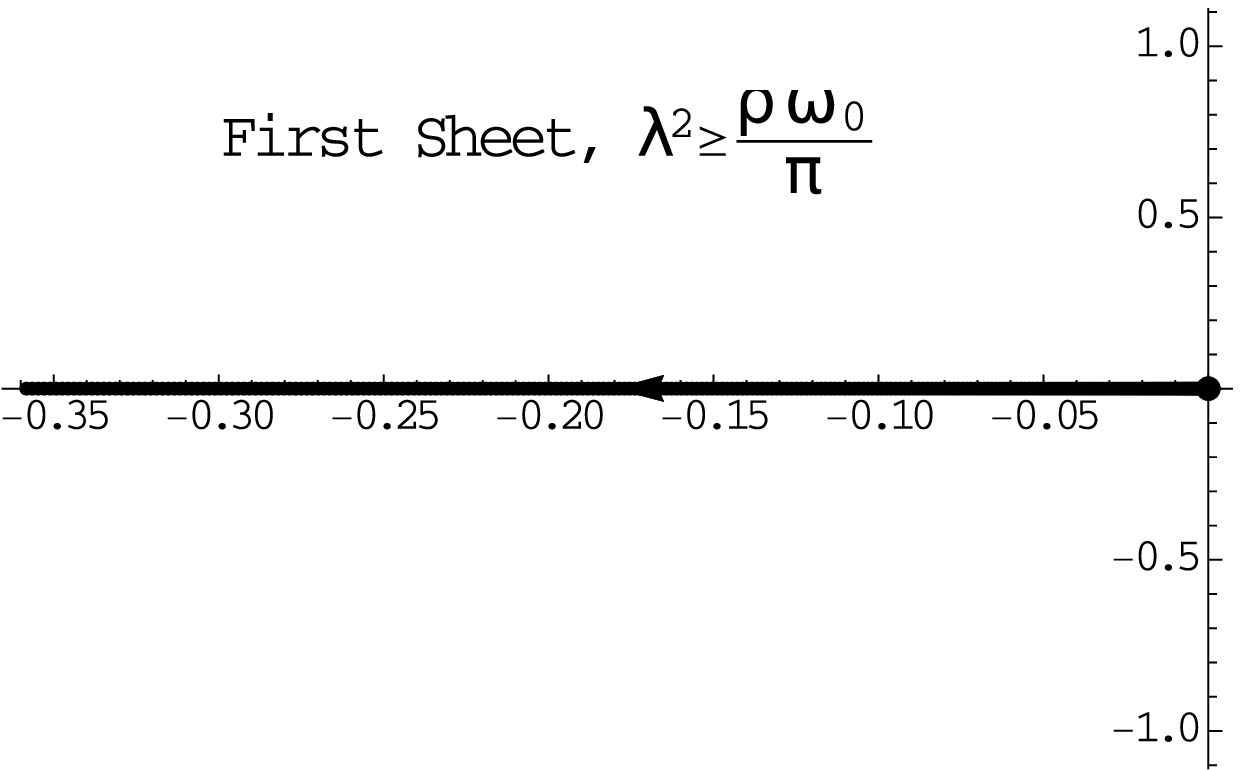}
\caption{The pole trajectories as $\lambda$ increases from $0$ when
$\omega_0<\frac 1 3 \rho^2$. We choose $\rho^2=3.1,\omega_0=1$. The
three second sheet poles merge and then separate. Two become a pair of
resonance poles and the other virtual state moves up to the first sheet.}
\label{fig:double}
\end{figure}

If  the relation between $\omega_0$ and $\rho$ in above condition (\ref{eq:triple-cond}) is not satisfied, one possibility is that the
two resonance poles can merge at a point different from the virtual
state when $\omega_0<\frac{\rho^2}3$.
In this case, we set the three pole positions as $u_{1,2}=-id$ and
$u_3=-id_0$, and the solution can then be obtained
\begin{align}
d=\frac 1 3\left(\rho\pm
\Big({\rho^2}-3\omega_0\Big)^{1/2}\right),
\quad
d_0=\frac 1 3\left(\rho\mp 2\Big({\rho^2}-3\omega_0\Big)^{1/2}\right)
\label{eq:double-solu}
\end{align}
There are two solutions which means there  are two points where either
the two
resonance poles come together to one point on the real axis or separate from one
point on the real axis.  The whole picture is as follows: The original two resonance poles merge first
and then separate on the negative real axis becoming two virtual
states. One virtual state moves down and
meets the original virtual state and then they separate into the complex
plane, becoming a pair of resonance poles. The other virtual state
moves through 
the threshold at the origin to the first sheet and becomes a
bound state. The bound state then moves down towards negative infinity
(see Fig. \ref{fig:double}).

Similar to the third-order pole, the second-order pole is a degenerate
of two states: the first state is the same as the ordinary virtual state
(\ref{eq:Virtual-state-right}), and  the second comes from the
second-order pole of the continuum state
\begin{align}
| z^\mp_{v2}\rangle=-N^\mp_{v2}
\lambda\int_0^\infty\frac{f(\omega)}{([z_v-\omega]_{\pm})^2}|\omega\rangle\mathrm{d}\omega
,
\quad \langle \tilde z^\pm_{v2}|= \langle  z^\mp_{v2}| \,,
\end{align}
The normalizations $N^-_v=N^-_{v_2}=(2/\eta'')^{1/2}$ and
$N^+_v=N^+_{v_2}=N^{-*}_v$ are chosen such that  $\langle \tilde
z_{v2}^\pm|z_{v}^\pm\rangle=1$. 
The second-order resonance poles were also 
found in a kind of one-dimensional double barrier potential~\cite{Mondragon:2000}
and in Friedrichs model using  another form factor in~\cite{Antoniou:2003}. Higher-order
resonance poles are found to represent the degenerate resonances in~\cite{Mondragon:1993}
 and are formulated in RHS
language in~\cite{Bohm:1997,Gadella:1997}.

\begin{figure}
\includegraphics[height=4cm]{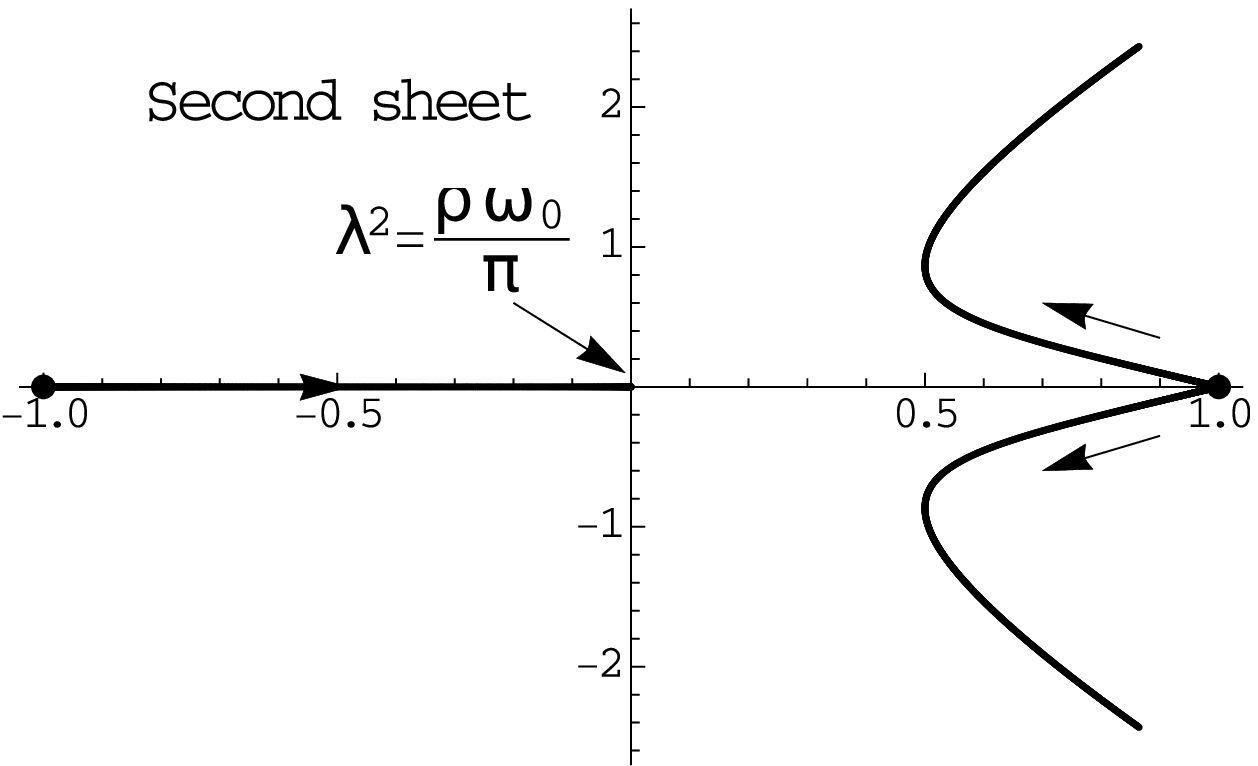}
\includegraphics[height=4cm]{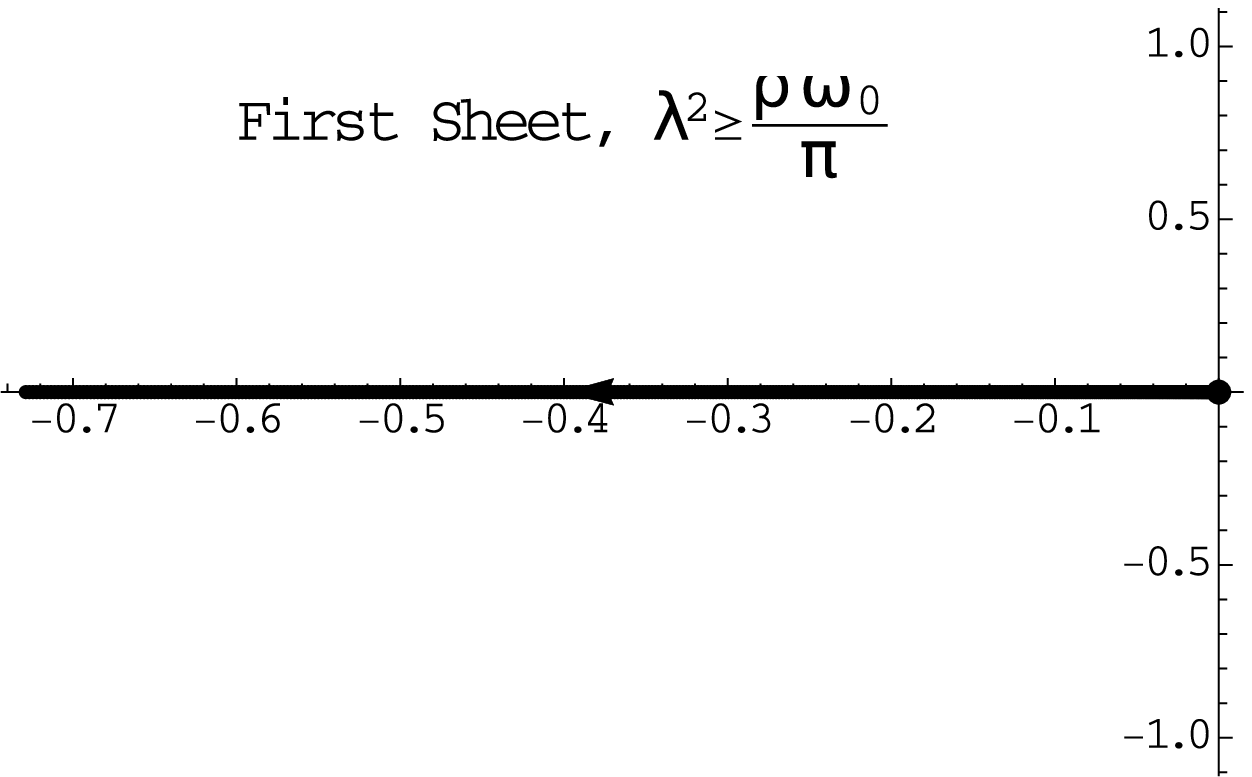}
\caption{The pole trajectories as $\lambda$ increases from $0$ when
$\omega_0>\frac 1 3 \rho^2$. We choose $\rho^2=1,\omega_0=1$. The
three second-sheet poles do not merge. The virtual-state pole moves up
to the first sheet for large $\lambda$.}
\label{fig:nonintersect}
\end{figure}

The other possibility is that the two resonance poles do not meet and
just run away towards infinity. The condition is $\omega_0>\rho^2/3$.
The virtual state in second sheet just moves to the threshold and comes
up to the first sheet, turning into a bound state, and then moves down toward
negative infinity ( see Fig. \ref{fig:nonintersect} ).

\begin{figure}
\includegraphics[height=4cm]{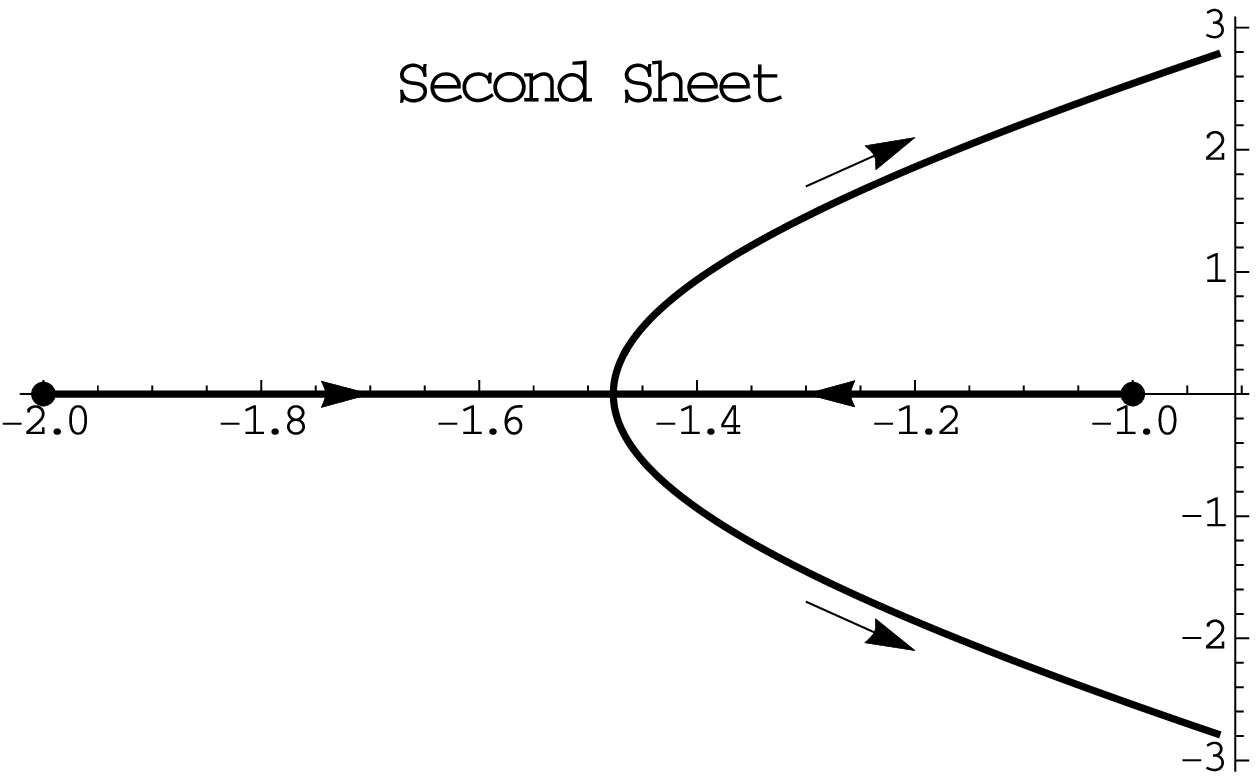}
\includegraphics[height=4cm]{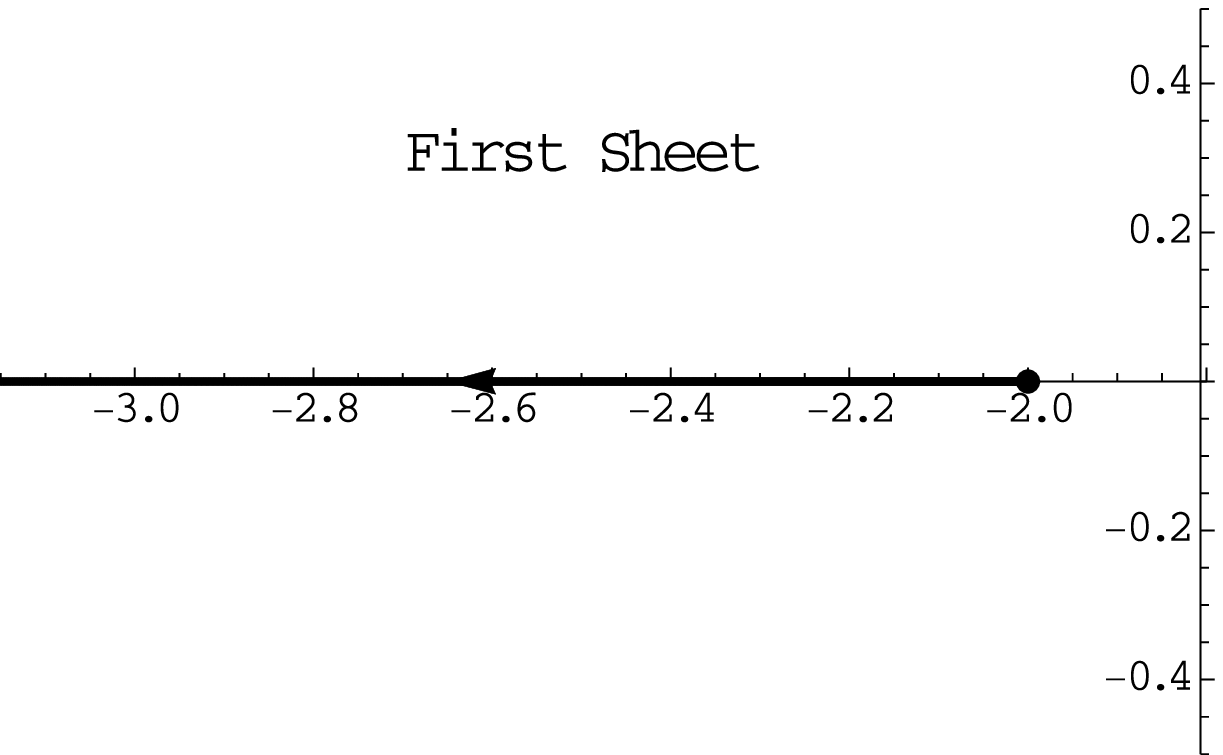}
\caption{The pole trajectories as $\lambda$ increases from $0$ when
$\omega_0<0$. $\rho^2=1,\omega_0=-2$. The two virtual poles merge
and then separate to a pair of resonance poles. The bound-state pole
moves to negative infinity.}
\label{fig:bound}
\end{figure}

{\it Case 3.} $\omega_0<0$. There is always a bound-state pole on the first
Riemann
sheet. $\eta=0$ gives
\begin{align*}
z_0=\omega_0+\lambda^2\int_0^\infty\frac{|f(\omega)|^2}{z_0-\omega}\mathrm{d}\omega
\end{align*}
Since the integral is
always negative,  $z_0$ is always smaller than $\omega_0$ which means the bound state
must move down from the original one. 
In our specific model, the bound state generated from the discrete
state on
the first sheet will always move down towards negative infinity.
For small $\lambda$, on the second sheet, besides the virtual pole
originating from the pole of the form factor, there is another virtual
state pole which is also generated from the original discrete state. 
We will give a general
argument of the existence of this virtual pole in the next section.
These two
virtual poles come together as $\lambda$ increases and merge to be a
second-order pole. As in previous case, at the merging point, we set
the three solutions to be  $u_{1,2}=-id$ and
$u_3=-id_0$, and then Eq. (\ref{eq:double-solu}) can also be used here
to describe the merging point. From (\ref{eq:double-solu}), $d$ can be positive
or negative. The negative one corresponds to an
imaginary $\lambda$ which is not physical. So, the two 
virtual-state poles can only merge once and then separately move onto the
complex plane and become a pair of resonance poles.
(see Fig. \ref{fig:bound}).

\section{Further discussion about the existence of the virtual
states\label{sect:General}}
 We have  seen that there could be  second-sheet virtual-state poles
generated from  the form factor or from the discrete states. In this
section we will give a general argument for these states to exist.

\begin{figure}
\begin{center}
\begin{center}\includegraphics[width=7cm]{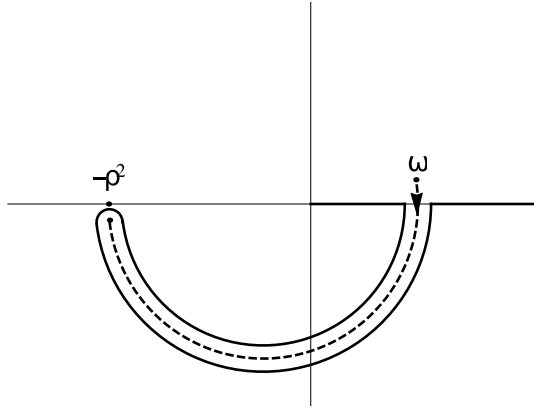}\end{center}
\end{center}
\caption{The integral contour is pinched by the two poles as the
$\omega$ analytically continued to the pole position of the form factor from the
first sheet to the second sheet.\label{fig:pinch}}
\end{figure}

\begin{figure}
\begin{center}
\includegraphics[height=4cm]{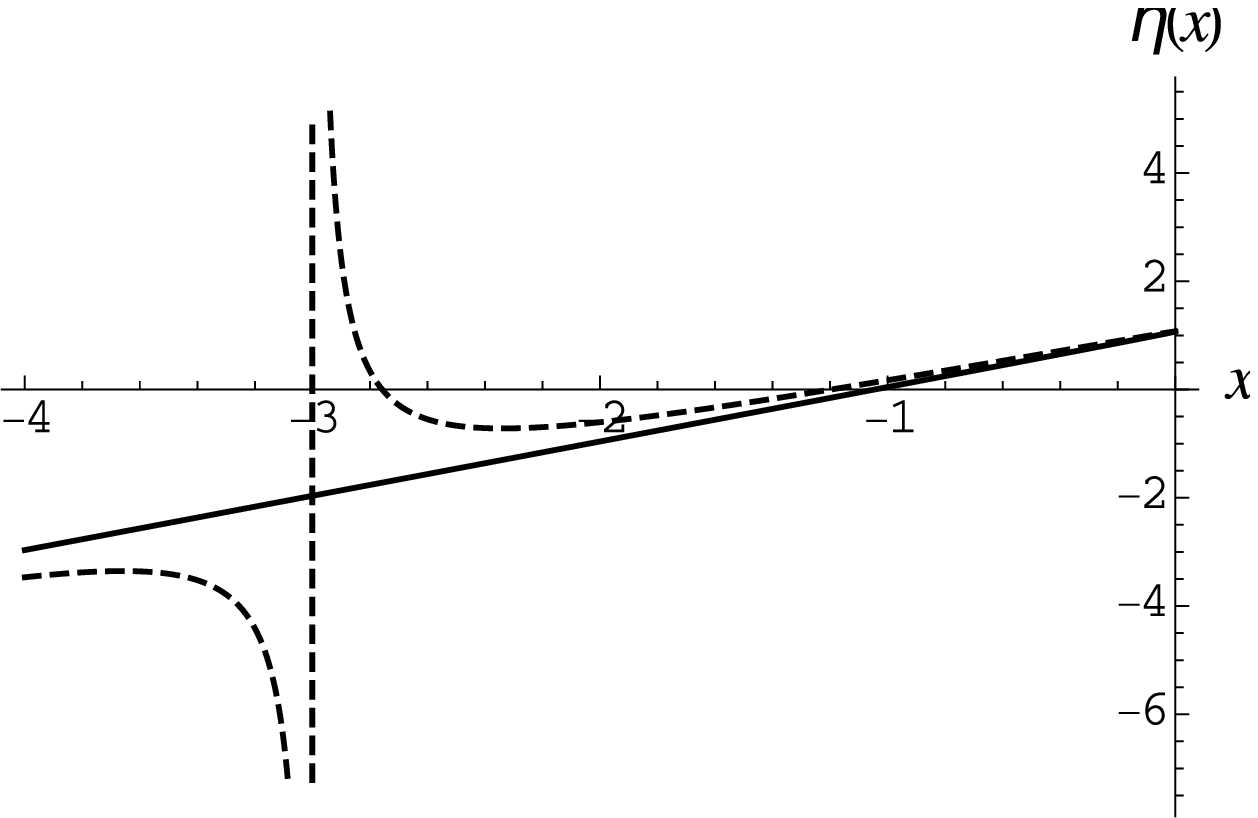}
\hspace{1cm}
\includegraphics[height=4cm]{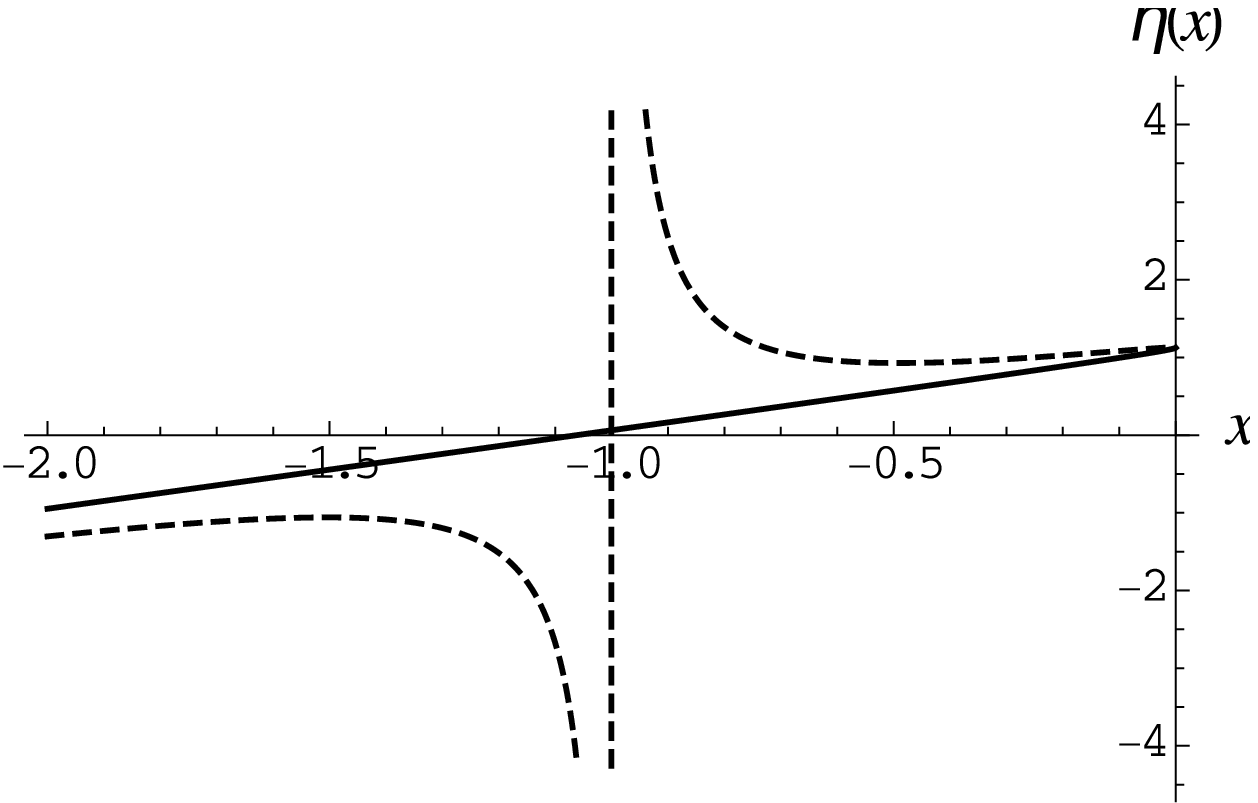}
\end{center}
\caption{Solid line and dashed line denote  $\eta(x)$ on the real axis of the
first and the second Riemann sheets, respectively. The left one :
$\lambda=0.2$, $\rho^2=3$ and $\omega_0=-1$, with two virtual states; The right one
$\lambda=0.2$, $\rho^2=1$ and $\omega_0=-1$, with a pair of resonance poles.
\label{fig:eta}}
\end{figure}

First we look at the virtual states generated from the form factor.
It is a general result that  whenever there is
a simple pole of the form factor, there is a
second sheet zero point of $\eta(\omega)$ near this position as the coupling constant is
turned on.  In \cite{Likhoded:1997}, the author noticed this phenomenon in
two special examples, but they did not provide a general reasoning for
this to happen.
Here, we provide a general argument as follows. Suppose the pole of the form
factor is at $\tilde \omega$ with $\tilde \omega\neq \omega_0$. The integral term in $\eta(\omega)$ in
(\ref{eq:eta-pm}) is analytic on the first sheet but has a
pole for $\eta(\omega)$ on the second sheet at the pole position of the form factor. This
pole appears because the analytic continuation of $\omega$  to the
second sheet to $\tilde \omega$ causes the integration contour to be
deformed and pinched by the two poles in the integrand as shown in
Fig. \ref{fig:pinch}.  
Then, $\eta$ on the  second sheet can be expressed as the first sheet $\eta$
plus a residue 
\begin{align}
\eta^{II}(\omega)=\eta^I(\omega)+2  \pi i\lambda^2\,
G^{II}(\omega)=\eta^I(\omega)-2 \lambda^2 \pi i\, G(\omega),
\end{align}
where the superscripts $I$ and $II$ denote the first sheet and  the
second sheet, respectively. The last equation can be deduced using the
Schwartz reflective relation for the real analytic function~\footnote{For $F(z)$ real on the negative axis and
complex on the positive real axis, $F(z)$  can be decomposed as $F(z)=R(z)+i
G(z)$, $R(z)={\rm Re} F(z)$, $G(z)={\rm Im }F(z)$,  for $z\in \mathbb
R_+$.  $R(z)$ and $G(z)$ are the analytic continuations of ${\rm Re}
F(z)$, ${\rm Im }F(z)$, respectively, from the positive axis. From
the Schwartz reflection
principle and the uniqueness of the analytic continuation,
$F^*(z)=F(z^*)$, $R^*(z)=R(z^*)$, $G^*(z)=-G(z^*)$ on the first sheet.
However, the Schwartz reflection principle also requires that the
continuation of $G(z)$ satisfies $G^*(z)=G^{II}(z^*)$, which means
$G(z)=-G^{II}(z)$.
 } and can also be understood as the analytic  continuation of $\eta(\omega)$ from the
lower rim of the cut up to the second sheet.
In our previous model,
\begin{align}
\eta^{II}(\omega)=\eta^I(\omega)
-2\lambda^2\pi i \frac{\sqrt \omega}{\omega+\rho^2}.
\end{align}
Since $\eta^I(\omega)$ is regular at the  $\tilde \omega$, the second
term gives the pole term at $\tilde \omega$ to $\eta^{II}$ which is
the pinch singularity.
Near this second-sheet pole of $\eta$,
$\eta^{II}(\omega)=0$ is of the form
\begin{align}
\frac{\lambda^2 c_1(\omega)}{\omega-\tilde \omega} =\omega -\omega_0\Rightarrow (
\omega -\omega_0)(\omega-\tilde \omega)=\lambda^2c_1(\omega),
\label{eq:eta-0}
\end{align}
where $c_1(\omega)$ is a function regular at $\tilde \omega$, and if
$\tilde \omega$ is real and below the threshold, $c_1$ is real.
 At $\lambda=0$, only $\omega_0$ is the
solution for the left equation.  However, as $\lambda$ is turned on, the zero point originating from $\tilde
\omega$ appears as illustrated in the left graph in Fig. \ref{fig:eta}. We can expand the solution with
respect to $\lambda^2$. Suppose the zero point of $\eta^{II}$ is at $\omega=\tilde
\omega+\lambda^2 \omega_2+O(\lambda^4)$, and above equation can be
recast as 
\begin{align}
-(\tilde \omega -\omega_0 )
\lambda^2 \omega_2 +O(\lambda^4)=\lambda^2c_1(\tilde
\omega)+O(\lambda^4)\Rightarrow  \omega_2=-\frac {c_1(\tilde \omega )}{\tilde
\omega-\omega_0}.
\end{align}
Thus, whenever the form factor has a pole, whether on the negative
real axis or on the complex plane, there will be a state generated
near the pole as  the coupling is turned on.

\begin{figure}
\includegraphics[width=8cm]{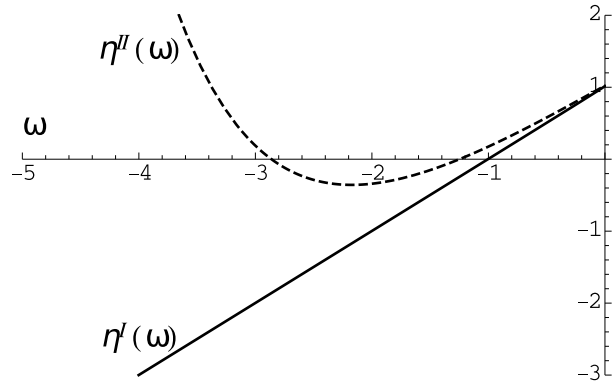}
\caption{$\eta(\omega)$ function on the first sheet  and second sheet
with form factor $G=\sqrt\omega e^{-\omega}$, $\omega_0=-1$, and
$\lambda=0.1$. \label{fig:Etazero}}
\end{figure}

In fact, a similar argument also applies to form factors such as
$G(\omega)=\sqrt{\omega}e^{-\omega}$.
In this case,
\begin{align}
\eta^{II}(\omega)=&\eta^I(\omega)-2  \pi i\, \lambda^2
G(\omega)=\omega-\omega_0-\lambda^2\int_0^\infty dx \frac
{G(x)}{(\omega-x)}
-2\lambda^2\pi i {\sqrt \omega e^{-\omega}}.
\\=&\omega-\omega_0+\lambda^2\int_0^\infty dx \frac
{G(x)}{(-\omega-x)}
+2\lambda^2\pi  {\sqrt {-\omega} e^{-\omega}}.
\end{align}
in which the integral in the last line is real and goes to zero as
$\omega\to -\infty$, while the last term goes to infinity at this
limit. At exactly $\lambda=0$, $\eta(\omega)=\omega-\omega_0$ and goes
to $-\infty$ as $\omega\to -\infty$.  As long as  $\lambda$ is turned
on, $\eta^{II}$ tends to $\infty$ as $\omega\to -\infty$. So for small
enough $\lambda$, there must be
one solution at large negative $\omega$ for the continuity of the
function. See Fig. \ref{fig:Etazero} for an
illustration.
 As $\lambda$ is turned on gradually, this
solution comes from  negative infinity. From this example we conclude
that if the form factor goes to $\infty$ as $\omega\to
-\infty$ faster than $-\omega$ without other singularities, there will be a virtual state
generated from the form factor from negative infinity as the coupling
is turned on.

Next we look at the virtual state generated from the discrete state.  In
fact, this is also a general result.
In general, whenever there is a
discrete state coupled to a continuum below the threshold, as soon as
the coupling is turned on, the original discrete state will be copied
on both the first sheet and the second sheet and be renormalized
separately. This is because before turning on the coupling
$\eta(z)=z-\omega_0$, ($\omega_0<0$ and $\omega_0\neq \tilde \omega)$,  and as we  turn on the coupling,
on both Riemann sheets, $\eta(z)$ has small real corrections on the
negative real axis near $\omega_0$, and the intersection between
$\eta(z)$ and the negative axis should also move away continuously
from $\omega_0$ on both sheets. See also the left figure in Fig. \ref{fig:eta}  for
illustration. 
We can also expand the solution around $\omega_0$ with respect to
$\lambda$, $\omega=\omega_0+\lambda^2
\omega_2+O(\lambda^4)$, and $\eta^{II}=0$ can also be expressed as
\begin{align}
0=\omega-\omega_0+\lambda^2 c_2(\omega)=\lambda^2 \omega_2+\lambda^2
 c_2(\omega_0)+O(\lambda^4) \Rightarrow \omega_2= -c_2(\omega_0)
\end{align}
where $c_2(\omega)$ is regular at $\omega_0$.
 Perturbation theory in quantum theory only concerns the bound states, and
the virtual state can only be reproduced by summing up the bubble
chain and solving the inverse propagator on the second Riemann sheet.
 One exception is when the pole position of the
form factor is the same as the energy for the discrete state
$\omega_0$, and $c_1(\omega_0)<0$, where $c_1$ is defined in
(\ref{eq:eta-0}).  The $\eta$ function is
illustrated in the right figure of Fig.\ref{fig:eta}.
In this accidental case, there are no solutions on the negative axis
but they will move to the complex plane and represent a pair of resonance
poles of the $S$ matrix as soon as $\lambda$ is turned on. This can be seen from
(\ref{eq:eta-0}), which becomes
\begin{align}
(\omega-\omega_0)^2=\lambda^2 c_1(\omega).
\end{align}
For small enough $\lambda$, we expand the solution with respect to
$\lambda$ as before, $\omega=\omega_0+\lambda \omega_1+O(\lambda^2)$, and
have
\begin{align}
\lambda^2 \omega_1^2+O(\lambda^3)=\lambda^2 c_1(\omega_0)+O(\lambda^3)
\end{align}
and for $c_1(\omega_0)<0$ , there can only be complex solutions.

\section{\label{sect:Complete} Completeness relation}

We have seen that besides the bound state and the continuum states,
there are  also resonant states and virtual states on the second sheet
of the Riemann sheet. When the discrete
state becomes resonance pole on the second sheet, only  the continuum
states form the complete set of bases and the completeness relation
can be expressed using the states defined in terms of
(\ref{eq:continuum-state-0}) as
\begin{align}
{\mathbf 1}=&\int_0^\infty d\omega |\Psi_+(\omega)\rangle \langle
\Psi_+(\omega) |. 
\end{align}

Nevertheless, in Ref. \cite{Prigogine:1991}, in order to solve the large
Poincar\'e problem, Petrosky, Prigogine, and Tasaki (PPT) proposed to
modify the continuum a little and then resonances can also appear in
the completeness relation.
In the spirit of PPT, the continuum states should be considered as a
complex functional with integral contour information encoded in it.
Two physical conditions, i.e., the decay of unstable state in the
future and the emission of the out-going wave, determine the rule for
choosing the integral contour.
In their derivation for the $|\Psi_+\rangle$ state, the physical condition
requires the 
pole position of the original discrete state to have a small $i\epsilon$ above the
path on the
real axis, and after the turning on of  the interaction, the pole
will move continuously down to the second sheet. The integral
path for the continuum state  should also be
continuously deformed to keep   the discrete state above the
integration path as in
Fig.\ref{fig:contour}.  Then one needs to include this path
information into the continuum state by using $\eta_d^+$, which can also be expressed as
\begin{align}
\frac 1{\eta_d^+(x)}\equiv\frac 1{\eta^+(x)}\frac {x-\tilde\omega_1+i
\gamma}{[x-\tilde\omega_1 +i\gamma]_+},
\label{eq:etap-d}
\end{align}
in which $\tilde \omega_1-i\gamma$ is the zero point of $\eta^+$ and
$[\dots]_+$ means deforming the contour as shown in Fig.
\ref{fig:contour}. Similarly, for $|\Psi_-\rangle$, $\omega_0$ is taken to be
a little below the real axis and  $\eta^-$ develops a zero above the
real axis on the second sheet. Then  the contour should be deformed
upward into the second sheet to keep the pole below the contour, and
one can include this contour information in $\eta_d^-=\frac
1{\eta^-(x)}\frac {x-\tilde\omega_1-i \gamma}{[x-\tilde\omega_1
-i\gamma]_-}$, where $[\dots]_-$ just denotes the deformation of the
contour from the lower rim of the real axis to upper the second sheet.
Now, the right continuum state can be expressed as
$\omega>0$,
\begin{align}
|\Psi_\pm(x)\rangle=&|x\rangle+\lambda\frac{f(x)}{\eta_d^\pm(x)}\Big[|1\rangle+\lambda\int_0^\infty\mathrm{d}\omega\frac{f(\omega)}{x-\omega\pm
i \epsilon}|\omega\rangle\Big]\,.
\label{eq:Continuum-state-right}
\end{align}
According to PPT's prescription, the left eigenstate with the same
eigenvalue would not need the deformation of the contour,
\begin{align}
\langle \tilde \Psi_\pm(x)|=&\langle
x|+\lambda\frac{f(x)}{\eta^\mp(x)}\Big[\langle
1|+\lambda\int_0^\infty\mathrm{d}\omega\frac{f(\omega)}{x-\omega\mp
i \epsilon}\langle\omega|\Big]\,.
\label{eq:Continuum-state-left}
\end{align}
The orthogonal relation for continuum states, $\langle \tilde \Psi_\pm
(x')|\Psi_\pm(x)\rangle=\delta(x-x')$, still holds. With these
definitions, the completeness relation can also be expressed in terms
of the continuum and the resonant state,
\begin{align}
{\mathbf 1}=&\int_0^\infty d\omega |\Psi_+(\omega)\rangle \langle
\tilde\Psi_+(\omega) | + |z_R\rangle \langle\tilde z_{R}|.
\end{align}

\begin{figure}
\begin{center}
\includegraphics[height=4cm]{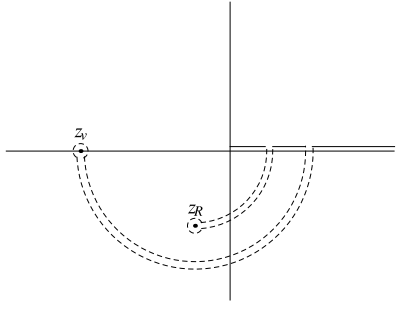}
\end{center}
\caption{The contour for $\eta_d^+$ in the case with one virtual pole
and a pair of resonance poles. \label{fig:Contour-V-R}}
\end{figure}

We can generalize these kinds of definitions to the cases with virtual
states. Since the virtual states and resonances may transform to 
each other as the coupling changes, one should not treat them differently. So, whenever there
are virtual states, it should also modify  the integral path for the
continuum states, and in the definition of $\eta_d^\pm$ in Eq.
(\ref{eq:etap-d}), all the second-sheet poles on the real axis and
the lower half-plane should be included, which means the integral path
is chosen as in Fig.~\ref{fig:Contour-V-R}.
With these definitions of the continuum, the generalized completeness
relation can be generalized here to  {\it Case 1}, including the virtual state from
the form factor
\begin{align}
{\mathbf 1}=&\int_0^\infty d\omega |\Psi_+(\omega)\rangle \langle
\tilde\Psi_+(\omega) | + |z_R\rangle \langle\tilde
z_{R}|+|z^+_v\rangle \langle\tilde z^+_{v}|.
\end{align}
In {\it Case 3}, where the discrete state becomes a bound state and a
virtual state for small $\lambda$, the generalized completeness relation should also include
the two virtual states
\begin{align}
{\mathbf 1}=&\int_0^\infty d\omega |\Psi_+(\omega)\rangle \langle
\tilde\Psi_+(\omega) | +\sum_{i=1}^2 |z^+_{vi}\rangle \langle\tilde z^+_{vi}|.
\end{align}

In general, as long as the contour for the continuum states goes around all the
virtual-state poles and the resonance poles on the lower second
sheet, these states also enter into the generalized completeness
relation along with
the bound states on the first sheet.

Above completeness relations only apply to simple poles. We have also
seen that there are second-order poles or third-order poles in our
example. One would wonder what role these states play in the
completeness relation.
In fact, for simplicity, if there is only one second-order virtual-state pole and no other
resonance or bound state, the generalized completeness relation is
\begin{align}
{\mathbf 1}=&\int_0^\infty d\omega |\Psi_+(\omega)\rangle \langle
\tilde\Psi_+(\omega) | + |z^+_v\rangle \langle\tilde z^+_{v2}|+ |z^+_{v2}\rangle
\langle\tilde z^+_{v}|\,,
\\
=&\int_0^\infty d\omega |\Psi_-(\omega)\rangle \langle
\tilde\Psi_-(\omega) | + |z^-_v\rangle \langle\tilde z^-_{v2}|+ |z^-_{v2}\rangle
\langle\tilde z^-_{v}|\,.
\end{align}
This can also be applied to the cases with complex double resonance
poles on the complex plane by changing the virtual states to the
corresponding resonant states. If there are more resonances or bound
states, they must also included in the sum as above. Similar
completeness relation in a special potential model in a different
context was
also found in~\cite{Hernandez:2003}.

This completeness relation can be generalized to the $n$th-order
virtual-state pole or resonance pole. In these cases, there are $n$
groups of  Gamow states for an $n$th-order pole,  which can be
represented as~\cite{Gadella:1997}
\begin {align}
|z^{(1)}\rangle=&N
\Big(|1\rangle+\lambda\int_0^\infty\frac{f(\omega)}{[z-\omega]_+}|\omega\rangle\mathrm{d}\omega\Big)\,,
\\
\langle \tilde z^{(1)}|=&N
\Big(\langle1|+\lambda\int_0^\infty\frac{f(\omega)}{[z-\omega]_+}\langle\omega|\mathrm{d}\omega\Big)\,,
\\
|z^{(n)}\rangle=&N(-1)^{n-1}
\lambda\int_0^\infty\mathrm{d}\omega\frac{f(\omega)}{([z-\omega]_+)^n}|\omega\rangle\,,\quad
\text{for } n\ge 2\,,
\\
\langle\tilde z^{(n)}|=&N(-1)^{n-1}
\lambda\int_0^\infty\mathrm{d}\omega\frac{f(\omega)}{([z-\omega]_+)^n}\langle\omega|\,,\quad
\text{for } n\ge 2\,,
\end{align}
$N=(\frac{n!}{\eta^{(n)}(z)})^{1/2}=\Big((-)^{n-1}\frac{\lambda^2} {n!} \int d
\omega \frac{|f(\omega)|^2}{([z-\omega]_+)^{n+1}}\Big)^{-1/2}$ is chosen such that $\langle
\tilde z^{(r)}|z^{(n-r+1)}\rangle =1$ and $z$ is the pole position either for
resonances or virtual states. These are the states continued from the
upper first sheet to the lower second sheet and there are also the
other set of states continued from the lower first sheet to the upper
second sheet. Suppose there are no other poles except the $n$th-order pole, 
the completeness relation can be expressed as
\begin{align}
{\mathbf 1}=&\int_0^\infty d\omega |\Psi_+(\omega)\rangle \langle
\tilde\Psi_+(\omega) | + \sum_{r=1}^{n}|z^{(r)}\rangle \langle\tilde
z^{(n-r+1)}|,
\end{align}
which is proved in the Appendix.
In general, if there is  more than one pole on the lower half-plane
and the negative axis of the second sheet at positions $z_{j}$,
($j=1,\dots,m_{II}$,) and also other simple bound-state poles
$z_{bi}$, $(i=1\dots,m_b)$ on the negative real axis of the first
sheet, with the continuum states modified accordingly, the most
general completeness relation can be written down as
\begin{align}
{\mathbf 1}=&\sum_i^{m_b} |z_{bi}\rangle\langle z_{bi}|+ \int_0^\infty d\omega |\Psi_+(\omega)\rangle \langle
\tilde\Psi_+(\omega) | + \sum_{j}^{m_{II}}\sum_{r=1}^{n_j}|z_j^{(r)}\rangle \langle\tilde
z_j^{(n_j-r+1)}|\,,
\end{align}
where $n_j$ is the order of the $j$th pole.

\section{\label{sect:Conclude} Conclusion and discussion}
In this paper, we first thoroughly studied the  Friedrichs model with an
integrable example form factor.  As discovered 
in~\cite{Likhoded:1997}, the form factor introduces extra second-sheet
poles besides the resonance poles originating from the discrete state.
We give a general argument that each pole of the form factor may introduce an
extra state on the second sheet. When the discrete state is below the
threshold, besides the bound state, there is also a virtual-state pole
arising from the discrete state for small coupling. We also give a
general argument for the existence of such a virtual-state pole.
This pole is similar to the shadow pole discussed in~\cite{Eden:1964zz} in
S-matrix theory.  We also generalize  PPT's prescription for the continuum states and
give the completeness relations including all the states on the
second Riemann sheet, i.e. all the virtual states and resonant
states arise both 
from the discrete states and from the form factor.  
We also find that for larger
coupling, the resonances that arise from the discrete state can meet on the
negative real axis to form a double pole  or meet with the other virtual
state to form a triple pole. We have
also discussed  the generalized completeness relations including these higher-order
resonances.

Note that in all the cases of the example, we have seen that the
original discrete state is
doubled as soon as the coupling is turned on. In {\it Case 1}, when
$\omega_0>0$, it becomes a pair of resonance
poles and in {\it Case 3} when $\omega_0<0$, it
becomes a bound state and a virtual state. In general, the number of the poles arising from the discrete state is
always doubled whenever a new threshold is opened. The poles on different
sheets with the same origin are called shadow poles by Eden and Taylor~\cite{Eden:1964zz}. 
However, the number of the  poles arising from the form factor is not
doubled when the first threshold is opened. This is the difference
between these two kinds of poles. The poles from the form factors can
be regarded as dynamically generated states. In fact, in the
dispersive analysis of the low-lying $0^{+}$ resonances, $\sigma$,
$\kappa$ are found to be just this kind of
resonance~\cite{Zhou:2010ra}.  In this paper, we only studied one
opened channel. The number of the poles from the discrete state is
only doubled once. When more channels are included, the number of the
poles may be doubled more than once.  In this sense, one original
discrete state may generate more resonant states to be observed in
the experiment. This phenomenon may already be observed in
the low-energy $0^+$ resonances~\cite{Zhou:2010ra}.

\begin{acknowledgments}
This work is supported by the China National Natural
Science Foundation under Contracts No.  11105138, No. 11575177 and
No. 11235010. Z.Z. also would like to thank the Project Sponsored by the Scientific Research Foundation for the Returned Overseas Chinese Scholars, State Education Ministry.
\end{acknowledgments}

\appendix

\section{Proof of the completeness relation for higher-order poles}
We assume that there is only one $n$th-order pole ($n\ge 2$) on the lower second
sheet. The completion relation is
\begin{align}
{\mathbf 1}=&\int_0^\infty d\omega |\Psi_+(\omega)\rangle \langle
\tilde\Psi_+(\omega) | + \sum_{r=1}^{n}|z^{(r)}\rangle \langle\tilde
z^{(n-r+1)}|.
\end{align}
{\bf Proof:}
\begin{align}
\int d\omega |\Psi(\omega)\rangle \langle \tilde \Psi(\omega)|=&\int
|\omega\rangle \langle \omega| +\int d
\omega
\frac{\lambda^2 |f(\omega)|^2}{\eta_d^+(\omega)\eta_(\omega)}\left[
|1\rangle \langle 1|+\int _0^\infty d\omega'\frac
{\lambda f(\omega')}{\omega-\omega'+i\epsilon}|\omega'\rangle \langle 1|
\right.
\nonumber \\
&\left.+\int _0^\infty d\omega'\frac
{\lambda f(\omega')}{\omega-\omega'-i\epsilon}|1\rangle \langle \omega'|
+
\int _0^\infty d\omega'd\omega''\frac
{\lambda^2 |f(\omega')|^2}{(\omega-\omega'-i\epsilon)(\omega-\omega''-i\epsilon)}|\omega'\rangle
\langle \omega''|\right]
\nonumber
\\&+\int _0^\infty d\omega\frac
{\lambda f(\omega)}{\eta^-(\omega)}|\omega\rangle \langle 1|
+\int _0^\infty d\omega\frac
{\lambda f(\omega)}{\eta_d^+(\omega)}|1\rangle \langle \omega|
\nonumber
\\&
+\int _0^\infty d\omega\frac
{\lambda f(\omega)}{\eta^-(\omega)}\int _0^\infty d\omega'\frac
{\lambda f(\omega')}{\omega-\omega'-i\epsilon}|\omega\rangle \langle
\omega'|
\nonumber
\\&+\int _0^\infty d\omega\frac
{\lambda f(\omega)}{\eta_d^+(\omega)}\int _0^\infty d\omega'\frac
{\lambda f(\omega')}{\omega-\omega'+i\epsilon}|\omega'\rangle \langle
\omega|
\label{eq:complete-1}\end{align}
Similar to the derivation in the Appendix of~\cite{Prigogine:1991},
the integrals above can be worked out as follows:
\begin{align}
I_1=&\int d\omega\frac {\lambda^2
|f(\omega)|^2}{\eta_d^+(\omega)\eta^-(\omega)}=-\mathrm{Res}\frac
1 {\eta(\omega)}\Big|^+_{z}+1=1\,,
\\
I_2(\omega')=&\int d\omega\frac {\lambda^2
|f(\omega)|^2}{\eta_d^+(\omega)\eta^-(\omega)(\omega-\omega'+i\epsilon)}
=-\frac
1 {\eta^-(\omega')}+\frac
{(-1)^nn!}{\eta^{+(n)}(z)[(z-\omega)_+]^n}\,,
\label{eq:I2}\\
I_3(\omega')=&\int d\omega\frac {\lambda^2
|f(\omega)|^2}{\eta_d^+(\omega)\eta^-(\omega)(\omega-\omega'-i\epsilon)}
=-\frac
1 {\eta^+_d(\omega')}+\frac
{(-1)^nn!}{\eta^{+(n)}(z)[(z-\omega)_+]^n}\,,
\label{eq:I3}\\
I_4(\omega',\omega'')=&\int d\omega\frac {\lambda^2
|f(\omega)|^2}{\eta_d^+(\omega)\eta^-(\omega)(\omega-\omega'+i\epsilon)(\omega-\omega'-i\epsilon)}
\\=&\frac 1{\omega'-\omega''-i\epsilon}\Big[-\frac
1 {\eta^-(\omega')}-\frac
1 {\eta^+_d(\omega'')}\Big]+\frac
{(-1)^nn!}{\eta^{+(n)}(z)}\sum_{r=0}^{n-1}\frac1{[(z-\omega)_+]^{r+1}[(z-\omega)_+]^{n-r}}\,.
\label{eq:I4}
\end{align}
Inserting the above equations in (\ref{eq:complete-1}), the terms
corresponding to the first terms in (\ref{eq:I2}) , (\ref{eq:I3}), and
(\ref{eq:I4}) cancel the
last three lines in  (\ref{eq:complete-1}) and we have
\begin{align}
\int d\omega |\Psi(\omega)\rangle \langle \tilde \Psi(\omega)|=&\int
|\omega\rangle \langle \omega| +|1\rangle \langle 1|
\nonumber \\&+\frac {(-1)^nn!}{\eta^{+(n)}(z)}
\Big[
\int _0^\infty d\omega'\frac
{\lambda f(\omega')}{[(z-\omega')_+]^n}|\omega'\rangle \langle 1|
+\int _0^\infty d\omega'\frac
{\lambda f(\omega')}{[(z-\omega')_+]^n}|1\rangle \langle \omega'|
\nonumber \\&
+\sum_{r=0}^{n-1}\int _0^\infty d\omega'\frac
{\lambda^2
f(\omega')f(\omega'')}{[(z-\omega')_+]^{r+1}[(z-\omega'')_+]^{n-r}}|\omega'\rangle
\langle \omega''|\Big]
\nonumber \\=&\mathbf 1- \Big[|z^{(n)}\rangle \langle 1|+|1\rangle\langle \tilde  z^{(n)}|+\int _0^\infty d\omega'\frac
{\lambda f(\omega')}{(z-\omega')_+}|\omega'\rangle\langle  \tilde z^{(n)}|
\nonumber \\&+\int _0^\infty d\omega'\frac
{\lambda f(\omega')}{(z-\omega')_+}|z^{(n)}\rangle\langle
\omega'|+\sum_{r=1}^{n-2}|z^{(r+1)}\rangle\langle  \tilde z^{(n-r)}|\Big]
\nonumber \\=
&\mathbf 1-\sum_{r=1}^{n}|z^{(r)}\rangle\langle \tilde z^{(n-r+1)}|
\end{align}
and  the completeness relation.
\bibliographystyle{apsrev4-1}

\bibliography{Ref}

\end{document}